# A modal perspective on the transverse Anderson localization of light in disordered optical lattices

Salman Karbasi
*Department of Electrical Engineering and Computer Science, University of Wisconsin-Milwaukee, Milwaukee, WI 53211, USA*

Karl W. Koch
*Optical Physics and Networks Technology, Corning Incorporated, SP-AR-01-2, Sullivan Park, Corning, NY 14831, USA.*

Arash Mafi*
*Department of Electrical Engineering and Computer Science, University of Wisconsin-Milwaukee, Milwaukee, WI 53211, USA.*

We frame the transverse Anderson localization of light in a one-dimensional disordered optical lattice in the language of localized propagating eigenmodes. The modal analysis allows us to explore localization behavior of a disordered lattice independent of the properties of the external excitation. Various localization-related phenomena, such as the periodic revival of a propagating Anderson-localized beam are easily explained in modal language. We characterize the localization strength by the average width of the guided modes and carry out a detailed analysis of localization behavior as a function of the optical and geometrical parameters of the disordered lattice. We also show that in order to obtain a minimum average mode width, the average width of the individual random sites in the disordered lattice must be larger than the wavelength of the light by approximately a factor of two or more, and the optimum site width for the maximum localization depends on the design parameters of the disordered lattice.

## INTRODUCTION

Ever since it was introduced in 1958, Anderson localization has been a subject of fascination and intense research in different subfields of theoretical and applied physics [1, 2]. The absence of diffusive transport of waves in a highly disordered medium was originally suggested for electronic quantum wavefunctions in atomic lattices; however, it was later realized that the concept applies equally well to classical wave systems, such as electromagnetics and optics [3–7]. In particular, localization of light in disordered media has attracted considerable attention; however, it is difficult to satisfy the strong scattering conditions required to achieve Anderson localization for optical waves in three dimensions [5]. The required conditions are considerably relaxed in lower dimensional systems. In fact, in unbounded one- and two-dimensional (1D and 2D) systems, all states are localized, regardless of the strength of disorder [8].

In bounded 1D and 2D systems, the localization can be observed as long as the transverse physical dimension(s) (size) of the system is larger than the localization length, which is inversely correlated with the disorder strength [9]. In optics, it is possible to introduce the disorder in only one dimension or two dimensions transverse to the (longitudinal) direction of light propagation; the optical beam will be localized in the disordered transverse direction(s) but can freely propagate in the invariant, longitudinal direction, similar to an optical waveguide. Transverse Anderson localization of light was originally proposed by De Raedt et al. [10]. Over the past few years, there have been several reports on the experimental observation of transverse Anderson localization of light in 1D [11–15] and 2D [16–18] disordered systems. In particular, we recently reported the development of polymer and glass optical fibers with 2D transverse disorder, where the waveguiding mechanism is maintained by transverse Anderson localization [17–19]. Anderson-localized optical fibers have many interesting and unconventional properties; e.g., unlike conventional optical fibers that can only transport light in their core region, Anderson localized fibers can trap and guide light coupled to any point across their input facet [19, 20].

Transverse Anderson-localized optical systems can act as optical waveguides and may be studied as such. In particular, modal analysis that is often employed to characterize optical waveguides can be a useful tool in developing proper physical intuition and providing reliable answers to often puzzling questions on the propagation and localization behavior in transverse-disordered optical systems.

The utility of the modal description has been recognized in several publications in recent years [11–14, 21, 22]. In this article, we take the modal analysis of the transverse Anderson localization one step further and use it to perform a detailed study of the localization phenomenon in a disordered 1D optical lattice.

There are several important advantages to the modal description of transverse disordered optical waveguides over other methods, such as the beam propagation method [10, 16, 19]. First, the modal framework recasts the complex physics of localization in a language more familiar to the practitioners in optics. Second, it can provide computational advantages in certain cases. Last, but foremost, the modal description is superior because it relies solely on the physics of the disordered system and is independent of the properties of the external excitation. In non-modal analyses and optimizations of disordered systems, it is often impossible to disentangle the impact of disorder in the waveguide from that of the in-coupling field; therefore, there is always the risk that a specific observed phenomenon is an artifact of the specifics of the optical excitation, rather than the properties of the disordered waveguide; the modal framework is most suitable to address this concern.

As an example of the utility of the modal description of transverse Anderson localization, we consider the strong analogy between the guiding mechanism in conventional optical fibers and Anderson-localized optical fibers [17]. In a glass rod with a uniform refractive index distribution, the transverse ($x$-$y$) profiles of the $z$-propagating guided modes extend over the entire cross section of the rod; however, when a core is introduced as in a conventional optical fiber, some of the guided modes of the rod localize in the core region. The transverse localization of optical modes in the core of a conventional fiber can be regarded as a result of transverse interference of light that results in a localized standing wave primarily over the core region. Similarly, in an optical fiber without disorder, the transverse ($x$-$y$) profiles of the $z$-propagating guided modes extend over the entire profile; however, when disorder is introduced, the guided modes of the system localize in the transverse direction. The localization is due to coherent multiple scattering and interference of the transmitted and reflected waves over the random structures [23], in strong analogy with conventional optical fibers. The main difference is that unlike conventional optical fibers, localized modes completely cover the entire cross section of the disordered fiber; therefore, Anderson-localized fibers behave effectively like a multicore optical fiber [20].

In this article, in order to present our observations and arguments, we have selected a disordered 1D array of slab optical waveguides, where we explore transverse localization for TE polarized optical fields. Because of the statistical nature of Anderson localization, our analyses are based on conducting multiple realizations of the disordered structure, sampled over a statistically identical ensemble. Our modal analysis provides a clear description of the localization strength in terms of the relevant design parameters for disordered arrays of 1D slab optical waveguides, such as the refractive index difference and the width of the constituent materials. Our hope is that such a modal description can also benefit the design and optimization of waveguides and optical fibers that use transverse Anderson localization as their waveguiding mechanism. We note that most of our observations are quite general and should equally apply, at least qualitatively, to TM waves, as well as to 2D disordered systems. The modal analysis of 2D disordered systems is computationally more challenging and will be presented elsewhere in the future, in the context of Anderson-localized optical fibers [17].

## CONSTRUCTING THE REFRACTIVE INDEX PROFILES

The 1D disordered optical waveguide studied here is constructed by perturbing a 1D *periodic* photonic crystal slab waveguide [24] in a specific way, as described in the next paragraph. The refractive index profile of the *periodic* waveguide (ordered waveguide) is sketched in Fig. 1(left), where it consists of $N_s$ slabs of equal thickness $\bar{d}_s = \bar{m}_s\Lambda$. We have chosen $\Lambda = \lambda/20$ as a fixed length-scale in this article and will also use it as the universal mesh size for numerical simulations of the waveguides, using the finite element method (FEM). $\lambda$ is the wavelength of light and $\bar{m}_s$ is a non-negative integer that indicates the thickness of each slab in units of $\Lambda$. The slabs have alternating refractive index values of $n_t$ and $n_b$, and the lattice is terminated on both sides by a cladding region of refractive index $n_c$. The total thickness of the waveguiding structure is assumed to be $N_s \times \bar{d}_s = 4000\Lambda = 200\lambda$. The cladding thickness is assumed to be $500\Lambda$ on each side, in order to ensure that the guided modes decay sufficiently in the cladding before reaching the cladding boundary, for accurate evaluation of the modes and their propagation constants with the FEM solver. Therefore, the width of the entire structure, including the cladding region, is $5000\Lambda = 250\lambda$. In Figs. 2, 3, 6, and 7, we present our plots only over the $200\lambda$-wide region of the lattice waveguide, while in Fig. 4 the entire $250\lambda$-wide structure is shown.

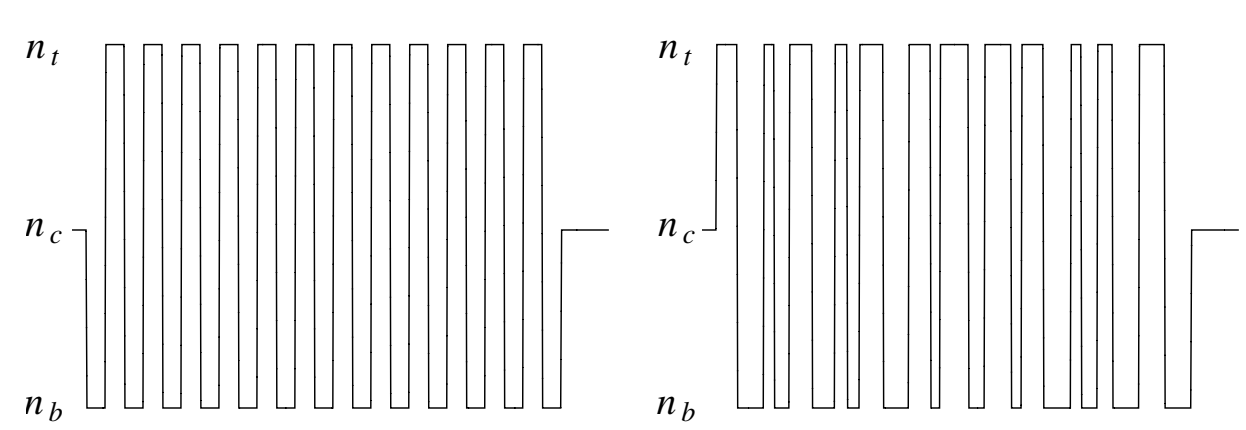

FIG. 1. Sample refractive index profiles of ordered (left) and disordered (right) slab waveguides are shown.

In order to construct the refractive index profile of the 1D disordered optical waveguide that is shown in Fig. 1(right), we perturb the thickness of each slab ($d_s = m_s\Lambda$) around a mean value ($\bar{d}_s = \bar{m}_s\Lambda$) by ($\delta_s = \eta_s\Lambda$). We assume that the perturbation $\eta_s$ is chosen from a zero-mean uniform discrete integer-valued distribution in the interval $[-\Delta_\eta, \Delta_\eta]$. For example, by choosing $\bar{m}_s = 20$ and $\Delta_\eta = 2$, we assume the thickness of each waveguide is randomly selected from $d_s = 18\Lambda$, $19\Lambda$, $20\Lambda$, $21\Lambda$, or $22\Lambda$, with equal probabilities. We note that in order to observe transverse Anderson localization, the disorder can be introduced in many different ways into the waveguide, such as diagonal [1] or off-diagonal disorder [14, 25, 26]; our method is similar to that of De Raedt et al. [10] (although not exactly the same), which is also closely related to the recent implementations of disordered optical fibers of Refs. [17–19] and relies on a combination of diagonal and off-diagonal disorder. In the next sections, we solve for the guided modes of the ordered and disordered lattices for propagation in the direction perpendicular to the optical lattices, in order to explore transverse Anderson localization, using the FEM technique originally presented in Ref. [27], but modified to solve for the TE modes in 1D slab waveguides.

## MODE PROFILES IN ORDERED AND DISORDERED WAVEGUIDES

In Fig. 2, we plot four arbitrarily selected guided modes of an ordered waveguide, where we have assumed that $n_t = 1.50$, $n_c = n_b = 1.49$, and $\bar{m}_s = 20$ ($\Delta_\eta = 0$). These four modes belong to a large group of standard "extended" Bloch periodic guided modes supported by the ordered optical waveguide, which are modulated by the overall mode profile of the slab-waveguide with a total thickness of $200\lambda$. The total number of guided modes depends on the total thickness and the refractive index values of the slabs and cladding. The key point is that each mode of the periodic structure extends over the entire width ($200\lambda$) of the waveguide structure.

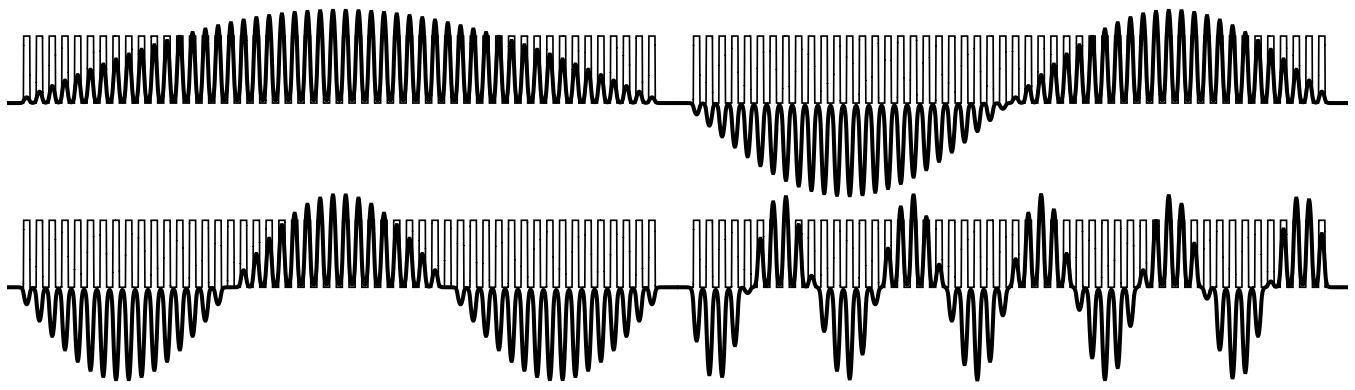

FIG. 2. Typical mode profiles of an ordered slab waveguide; each mode extends over the entire waveguide.

We now perturb the ordered array of Fig. 2 and show that the perturbation results in the localization of the guided modes. In Fig. 3, we plot several modes of disordered waveguides, where again we assume that $n_t = 1.5$, $n_c = n_b = 1.49$, and $\bar{m}_s = 20$. The top two plotted modes in Fig. 3 are selected among the guided modes of a single realization of a disordered waveguide with $\Delta_\eta = 3$, and the bottom two modes are selected among the guided modes of a single realization of a disordered waveguide with $\Delta_\eta = 10$. While there are variations in the width of the modes in each disordered waveguide, the mode profiles shown in Fig. 3 are typical and clearly illustrate that the disorder is behind the localization of the guided modes. Moreover, the degree of localization is increased when the disorder level is larger: the guided modes become narrower when $\Delta_\eta$ gets larger. This observation universally applies to *almost* all the guided modes of a disordered structure, as shown in Fig. 6. In bounded systems, the presence of the boundary affects the localization properties of some modes and can even result in delocalization [13, 19, 21, 28]; however, in the lattices studied here all the guided modes are well localized.

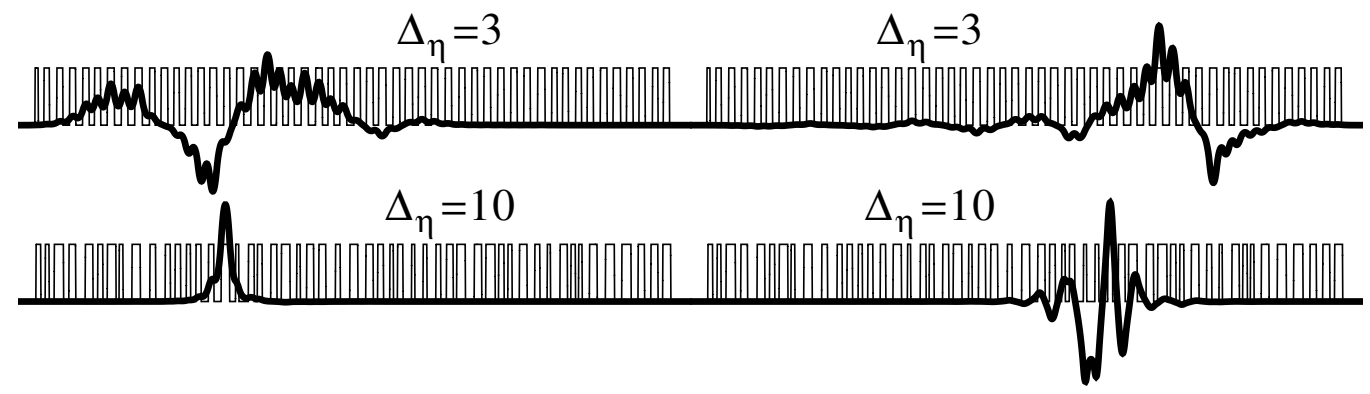

FIG. 3. Typical mode profiles of disordered slab waveguides for $\Delta_\eta = 3$ (lower disorder) and $\Delta_\eta = 10$ (higher disorder), where $\bar{m}_s = 20$. The modes are localized and the localization effect is stronger for higher disorder.

In order to visualize the propagation and localization of a beam coupled to a disordered waveguide with localized modes, we inject a Gaussian beam to the disordered waveguide and plot the intensity distribution of the guided beam as it propagates along the waveguide, shown in Fig. 4. The disordered waveguide is defined by $\bar{m}_s = 20$, $\Delta_\eta = 20$, $n_t = 1.50$, $n_b = 1.45$, and $n_c = 1.40$, and the Gaussian beam is characterized by the electric field distribution of the form $E(x) \propto \exp(-x^2/w^2)$ with $w = 2\lambda$ at the entrance, $x$ is the coordinate across of the waveguide. The center of the Gaussian beam is assumed to be in the middle of the waveguide. The excitation amplitudes of the modes are calculated using the overlap integrals of the in-coupling Gaussian beam with the guided modes of the structure [29]. The density plot of the optical intensity of the propagated beam shown in Fig. 4 is given by $I(x,z) = |E(x,z)|^2$, where the electric field $E(x,z)$ is propagated according to $E(x,z) = \sum_j A_j F_j(x) \exp(i\beta_j z)$. Here, $A_j$, $F_j$, and $\beta_j$ are the excitation amplitudes, transverse mode profiles, and propagation constants, respectively, associated with the mode number $j$, where the sum runs over all the guided modes of the lattice. In the single realization of the disordered waveguide shown in Fig. 4, 96% of the power in the Gaussian beam is coupled to the guided modes of the disordered waveguide; the remaining 4% is coupled to the radiation modes that quickly radiate away and are not shown in Fig. 4.

The Gaussian beam only couples efficiently to those guided modes that are localized near the center of the waveguide. Near the entrance, the guided excitation in the waveguide closely resembles the in-coupling Gaussian beam, although it is not exactly identical because some power is also coupled to the continuum of the radiation modes. Each excited mode propagates with a different phase velocity, determined by their individual propagation constants; therefore, the *detailed balance* between the excitation amplitudes of the guided modes that is responsible for the narrow excitation at the entrance is broken as the relative phases between the modes change as the beam propagates along the waveguide. As the beam propagates and the detailed balance is further broken, the beam expands; however, it can never expand beyond the distribution of its constituent modes and its expansion is eventually halted, as observed in previous studies of transverse Anderson localization. We note that the average localized width of the beam is likely to be larger than the typical width of its constituent modes, because there is a larger distribution of the central positions of the excited modes across the disordered waveguide.

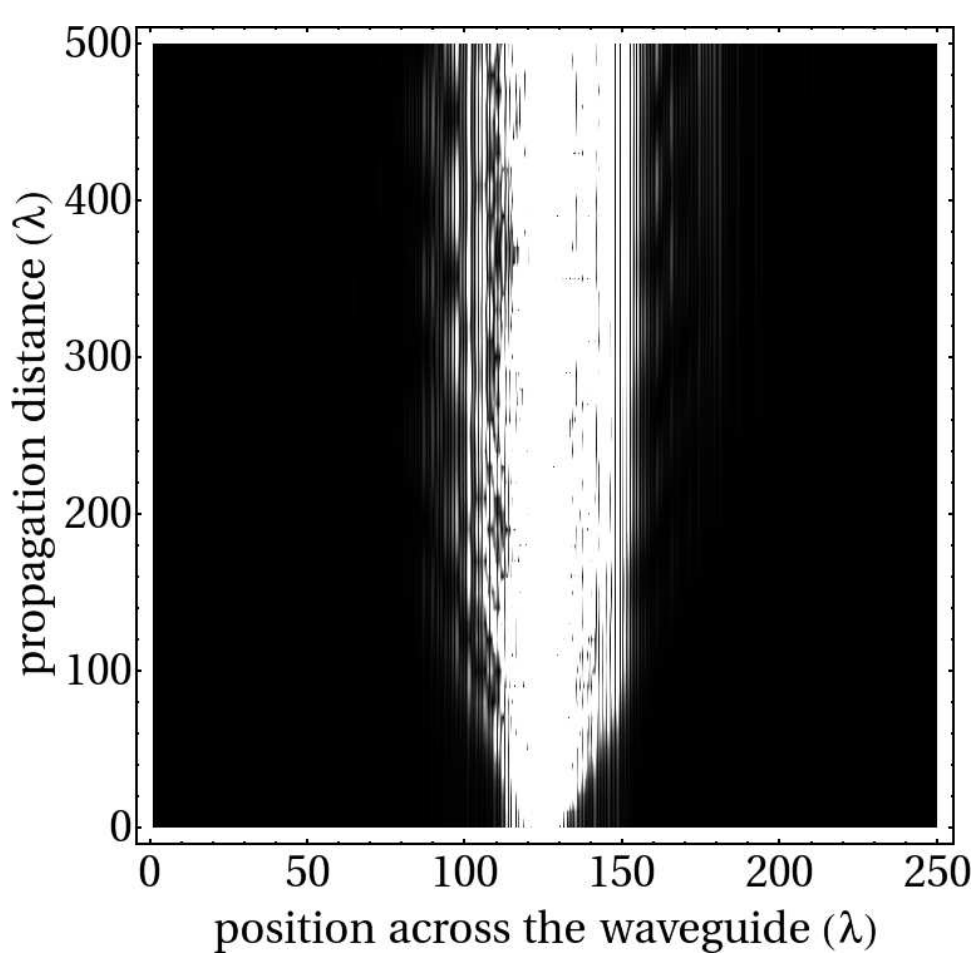

FIG. 4. A Gaussian beam is coupled to a disordered 1D waveguide. The intensity distribution shows that the beam goes through an initial expansion and eventually localizes to a relatively stable width.

The modal picture can also provide a convincing answer to another recurring question on disordered waveguides: what happens if the initial excitation is larger than the typical localization width of a disordered waveguide? Will a wide beam focus to the localization radius? The answer is that a wide beam excites a larger number of localized guided modes across the waveguide and the final localized intensity distribution will inevitably be larger, determined by the width of the excited constituents, as well as their distribution across the waveguide. We note that it is possible to obtain focusing in a typical randomized lattice such as in Fig. 4. This can be achieved simply by time reversing the propagation shown in Fig. 4 using phase conjugation; if the in-coupling beam is exactly equal to the complex conjugate of the field shown in Fig. 4 at $z = 500\lambda$, it will focus to the narrow Gaussian beam of $z = 0$ after propagating a distance of $500\lambda$ [30, 31]. For example, this may be accomplished by means of a very careful spatial wavefront shaping using a spatial light modulator.

Related to the point discussed above, because of the discreteness of the guided modes, the field distribution along the waveguide is periodic with the propagation distance and the beam will eventually refocus back to the initial distribution (self-imaging), if it propagates a long enough distance. The revival distance is related to the differences between the propagation constants in the excited guided-mode superposition. This effect is shown in Fig. 5, which is identical to Fig. 4 except the beam propagation is shown over a longer distance of $1500\lambda$, where the revival of the in-coupling field can be observed at the propagation distance of $\approx 1000\lambda$. The periodic evolution of the Anderson-localized beam has recently been observed experimentally by El-Dardiry, et al. [11]. We would like to note that this effect is well known in atomic physics, where the periodic revival of the quantum wavefunction has been reported and is rooted in the discreteness of the eigenstates of the system from which only a finite number are excited, given an initial field [32–34].

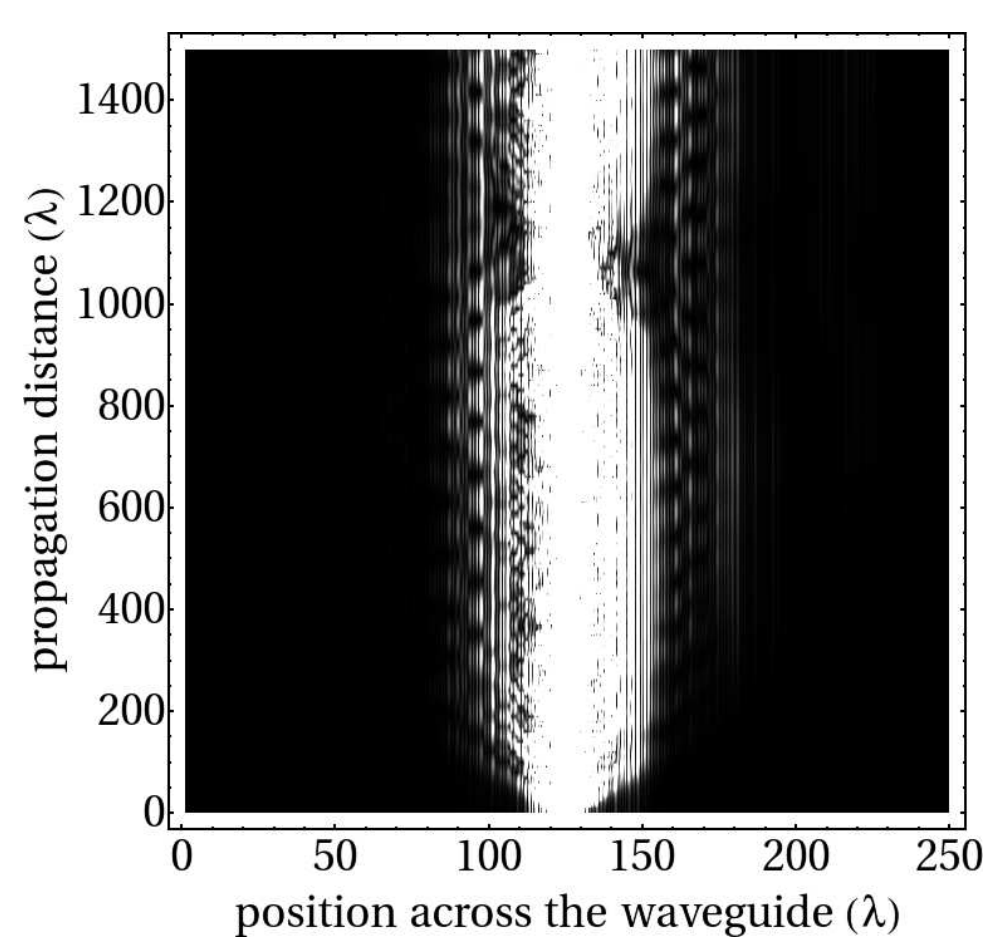

FIG. 5. Gaussian beam excitation of the lattice same as Fig. 4, but for a longer propagation distance to observe the revival of the in-coupling field at the propagation distance of $\approx 1000\lambda$.

We note that the random distribution of the propagation constants as well as the large number of excited modes may result in a very large self-imaging period; therefore, the typical picture of an initial expansion and an eventual localization can hold quite well as long as the propagation distance of the beam remains below the revival period.

## EFFECT OF THE INDEX DIFFERENCE AND LEVEL OF DISORDER ON THE DISTRIBUTION OF THE MODE WIDTHS

The distribution of the mode widths for all the guided modes of individual realizations of a disordered waveguide with $\bar{m}_s = 20$ are shown in Fig. 6 for $\Delta_\eta = 20$ (circles) and $\Delta_\eta = 5$ (triangles), where $n_t = 1.50$, and $n_c = n_b = 1.49$. Each circle (or triangle) represents a mode, where the horizontal coordinate is the "mean" position of the mode across the waveguide and the vertical coordinate is the width of the mode. The mean position and the width of each mode is calculated using the second moment method [35] based on the intensity profile of the mode $I(x)$. For each guided mode, we calculate the mode position ($\bar{x}$) using $\bar{x} = \int_\infty^\infty dx\ xI(x)$ and the mode width ($\sigma$) using $\sigma^2 = 2\int_\infty^\infty dx\ (x-\bar{x})^2 I(x)$, where the modes are normalized according to $\int_\infty^\infty dx\ I(x) = 1$. We emphasize that the vertical axis in Fig. 6 is in logarithmic scale; therefore, there is a substantial reduction in the typical values of the mode width as the disorder level is increased.

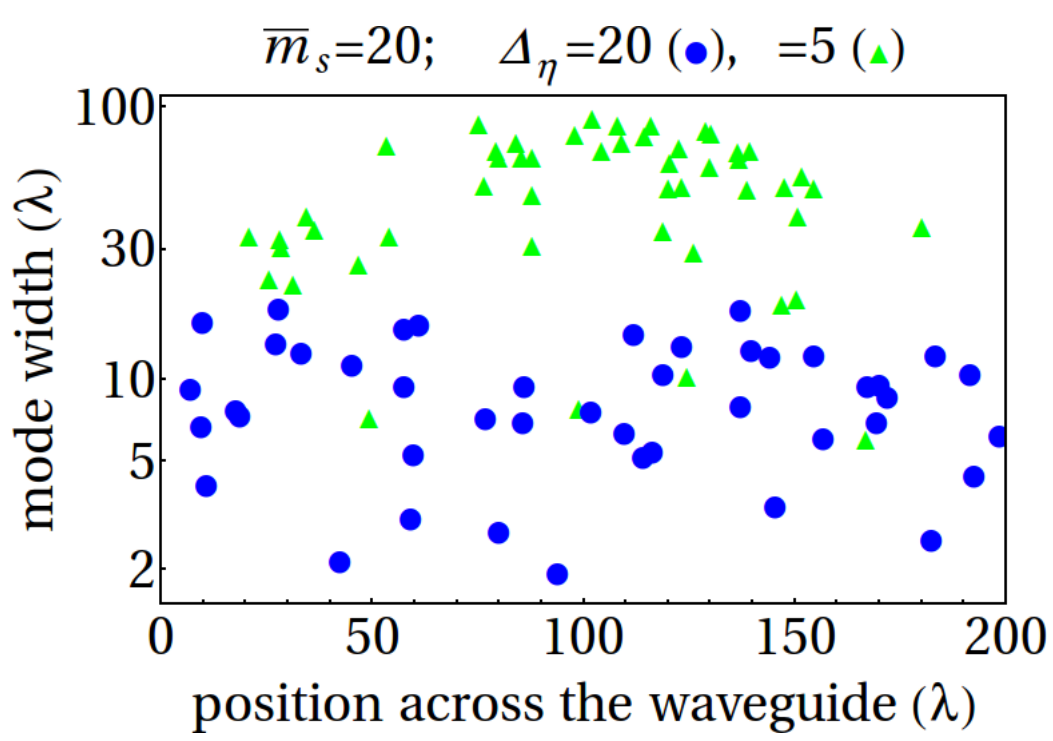

FIG. 6. The distribution of the mode widths for all the guided modes of a disordered waveguide for two different levels of disorder are compared. Each point represents the position (x-axis) and the width (y-axis) of the mode, where the circles and triangles represent $\Delta_\eta = 20$ and $\Delta_\eta = 5$, respectively. The modes of the waveguide with a higher level of disorder ($\Delta_\eta = 20$) are more localized.

We recently carried out a detailed study on the dependence of the localization radius in an Anderson-localized optical fiber on the index difference between the constituents of the disordered fiber [19]; here, we suggest that the reason behind the dependence of the localization radius on the index difference is that the guided modes of the disordered waveguides are more localized when the refractive index difference is increased. For our 1D disordered waveguides, we clearly observe that the width distribution of the localized guided modes depends strongly on the index difference of the alternating slabs $\Delta n = n_t - n_b$. In Fig. 7, we present the mode width distribution for all the guided modes of individual realizations of a disordered waveguide with $\bar{m}_s = 20$ and $\Delta_\eta = 10$ for $\Delta n = 0.1$ (circles) and $\Delta n = 0.01$ (triangles), where $n_t = 1.50$ and $n_c = n_b$; both cases have identical levels of disorder ($\Delta_\eta$). Fig. 7 confirms that the mode widths are substantially reduced as the index difference $\Delta n$ is increased.

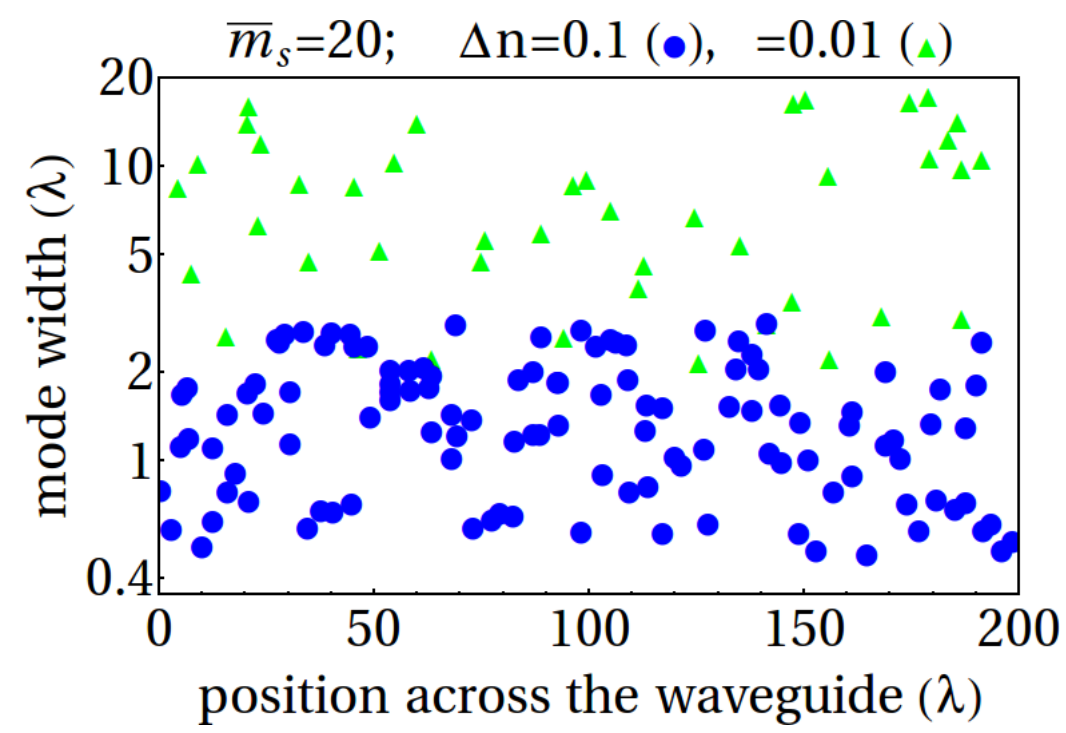

FIG. 7. The distribution of the mode widths for all the guided modes of a disordered waveguide for two different values of the index difference $\Delta n$ and identical levels of disorder $\Delta_\eta$ are compared. Each point represents the position (x-axis) and the width (y-axis) of the mode, where the circles and triangles represent $\Delta n = 0.1$ and $\Delta n = 0.01$, respectively. The modes of the waveguide with a larger index difference ($\Delta n = 0.1$) are more localized.

In order to visualize the width distribution of the guided modes in a disordered lattice, we consider the case of $\bar{m}_s = 40$ and $\Delta_\eta = 20$ and present the probability distribution of the mode widths in units of $\lambda$ for two separate cases of $\Delta n = 0.1$ and $\Delta n = 0.01$ in Fig. 8. Each probability distribution is presented in a histogram and is the result of averaging 100 independent random simulations. It can be clearly seen in Fig. 8 that $\Delta n = 0.1$ results in a much stronger localization than $\Delta n = 0.01$. Not only is the peak of the distribution for $\Delta n = 0.1$ located at a smaller value of mode width compared with $\Delta n = 0.01$, but also the standard deviation of the distribution is much smaller; this observation is related to our previously reported reduction of the standard deviation and sample-to-sample variation in Anderson-localized optical fibers and is rooted in stronger self-averaging behavior when the localization effect is stronger [17, 19, 23].

### Metric for the strength of the localization

In order to compare the localization strength of various disordered lattices, one needs to devise a metric that is universal (independent of the in-coupling beams) that can, at least, easily distinguish between two lattices with substantial differences in their localization strengths. Here, we use the average width of all the guided modes in the lattice. However, occasionally, there are a few modes that remain extended and that can bias the averaging process. In order to eliminate this undesired bias, we do not consider those modes whose width is larger than twice the median value of mode widths from the set of the guided modes; after this elimination, the resulting average comes out very close to the median. Because the localization process is stochastic, a more reliable metric is obtained if it is also averaged over many members of a statistically identical ensemble. Therefore, for each disordered lattice defined by a given set of $\bar{m}_s$, $\Delta_\eta$, and $\Delta n$, we also average the "mean beam width" over 100 different random realizations of the lattice. The final results are shown in Figs. 9, 10, 11, and 12 for different values of $\bar{m}_s$, $\Delta_\eta$, and $\Delta n$. We note that we could have kept more modes in our averaging process by only eliminating those modes whose width is larger than, e.g., three times the median value, but the resulting metric would not have changed the main conclusions presented in this article.

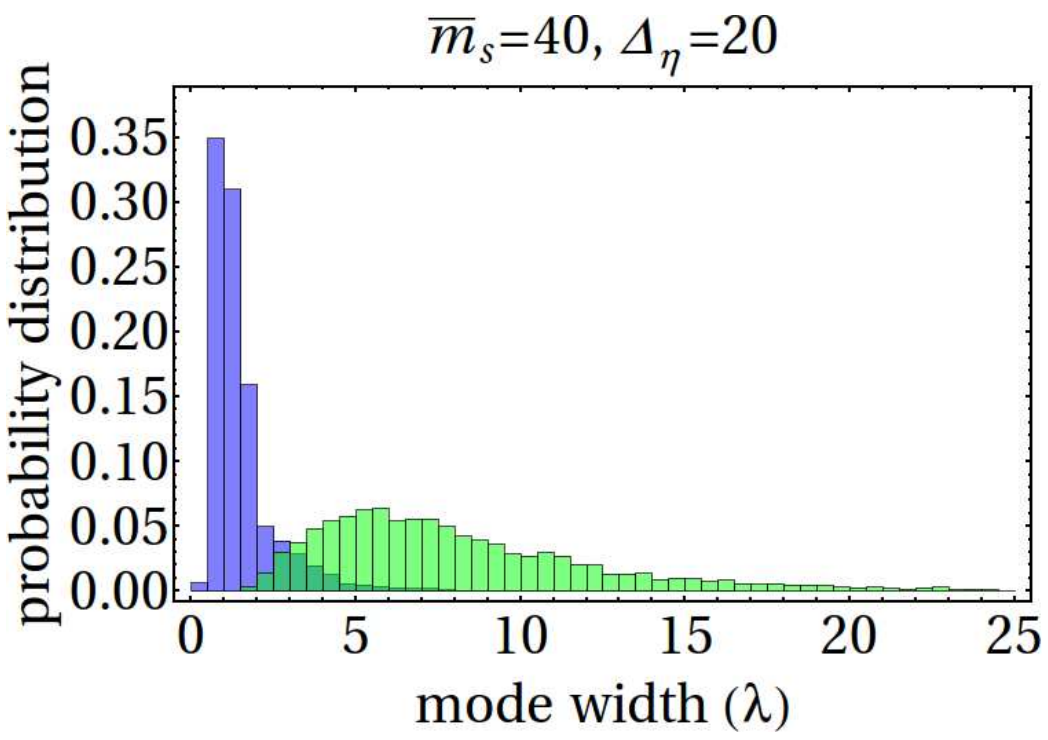

FIG. 8. The probability distribution of the mode widths for two separate cases of $\Delta n = 0.1$ and $\Delta n = 0.01$ in a disordered lattice defined by $\bar{m}_s = 40$ and $\Delta_\eta = 20$. Each histogram is the result of averaging over 100 independent random simulations.

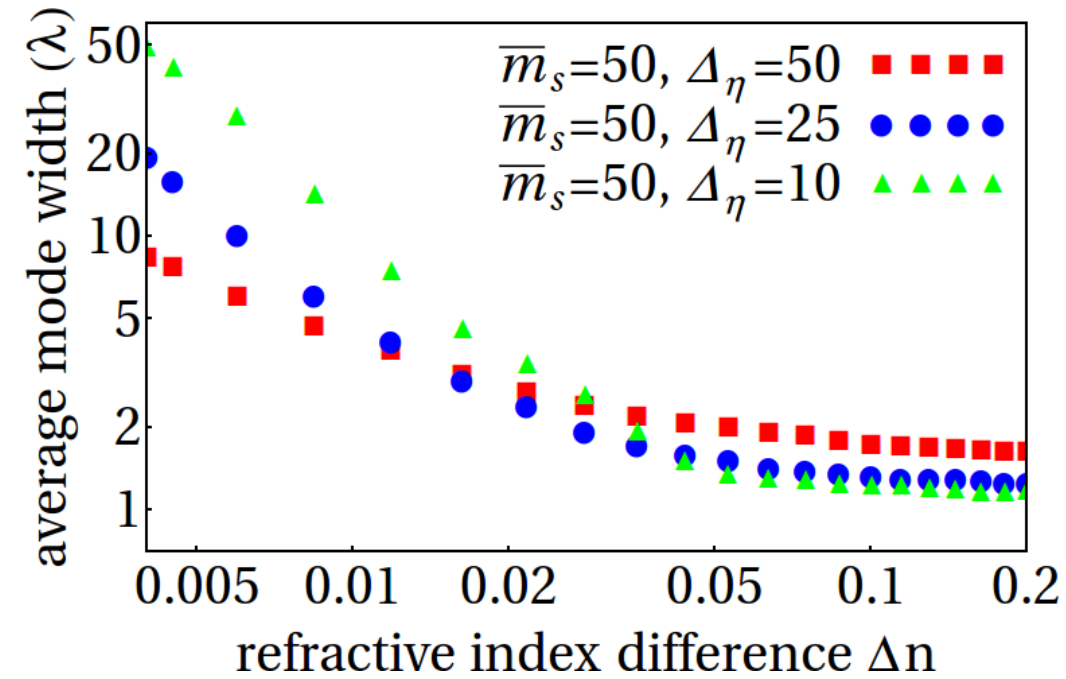

FIG. 9. The average mode widths of different disordered lattices are compared in units of $\lambda$ for $\bar{m}_s = 50$ (fixed mean slab width) as a function of $\Delta n$. The results are presented for three different values of disorder: $\Delta_\eta = 50$ (squares), $\Delta_\eta = 25$ (circles), and $\Delta_\eta = 10$ (triangles).

### Impact of the level of disorder and index difference

In Fig. 9, we fix the mean width of each slab by choosing $\bar{d}_s = 2.5\lambda$ ($\bar{m}_s = 50$) and plot the modes' average width (averaged over the guided modes of 100 random realizations of the lattice) as a function of $\Delta n$, for three different values of disorder: $\Delta_\eta = 50$ (squares), $\Delta_\eta = 25$ (circles), and $\Delta_\eta = 10$ (triangles). We note that the plot is in logarithmic scale for both axes to provide a clearer presentation.

For a fixed disorder level (fixed value of $\Delta_\eta$), the average mode width decreases as $\Delta n$ is increased. However, this decrease eventually levels off at around $\Delta n = 0.1$. This observation has important implications for the optimal design of disordered waveguides; e.g., the Anderson-localized optical fiber in Ref. [17] is designed with $\Delta n = 0.1$. However, we caution the reader on directly applying the results obtained in this 1D analysis to 2D disordered geometries.

We next compare the localization properties of lattices with different disorder levels in Fig. 9. A comparison between the average mode widths at a fixed value of $\Delta n$ (squares versus circles versus triangles) shows that at least for $\Delta n \lesssim 0.02$, larger disorder (larger $\Delta_\eta$) results in a smaller average mode width, which is also intuitively easy to comprehend. However, as the index difference $\Delta n$ is increased, the data related to different values of $\Delta_\eta$ cross over each other and it appears that at $\Delta n = 0.2$, the average mode width is smallest when $\Delta_\eta$ is smallest. While this is quite an interesting and notable behavior, it could just simply be an artifact of how we designed our metric, given that the average mode widths are not very different at $\Delta n = 0.2$ for different values of $\Delta_\eta$. We do not recommend using the exact values presented in Fig. 9 to make definitive judgments on the localization strengths of different lattices at large $\Delta n$; rather, we think that the main message of our results is that a large value of $\Delta n$ is more forgiving in terms of the disorder level in obtaining a sufficiently strong localization behavior.

### Impact of the mean slab width and index difference

We can now explore what happens when the disorder level $\Delta_\eta$ is fixed but other lattice parameters are changed. In Fig. 10, we fix the disorder level at $\Delta_\eta = 10$ and plot the average beam width as a function of $\Delta n$ for three different values of the mean slab width given by $\bar{m}_s = 10$ (squares), $\bar{m}_s = 30$ (circles), and $\bar{m}_s = 50$ (triangles).

The results show a mixed behavior and it is not possible to come up with a verdict that equally applies to the entire parameter space. However, from a practical design standpoint, it is remarkable that the case of $\bar{m}_s = 50$ (triangles) saturates to a small average mode width for values of index difference as low as $\Delta n = 0.05$; therefore, it is possible to obtain reasonably strong localization, even when the lattice constituents are as thick as $d_s = 2.5\lambda \pm 0.5\lambda$. Another remarkable point is that for large $\Delta n$, $\bar{m}_s = 30$ results in a lower average mode width than both $\bar{m}_s = 50$ and $\bar{m}_s = 10$, indicating that numerical optimizations are likely necessary to obtain the best localization behavior for device applications [19].

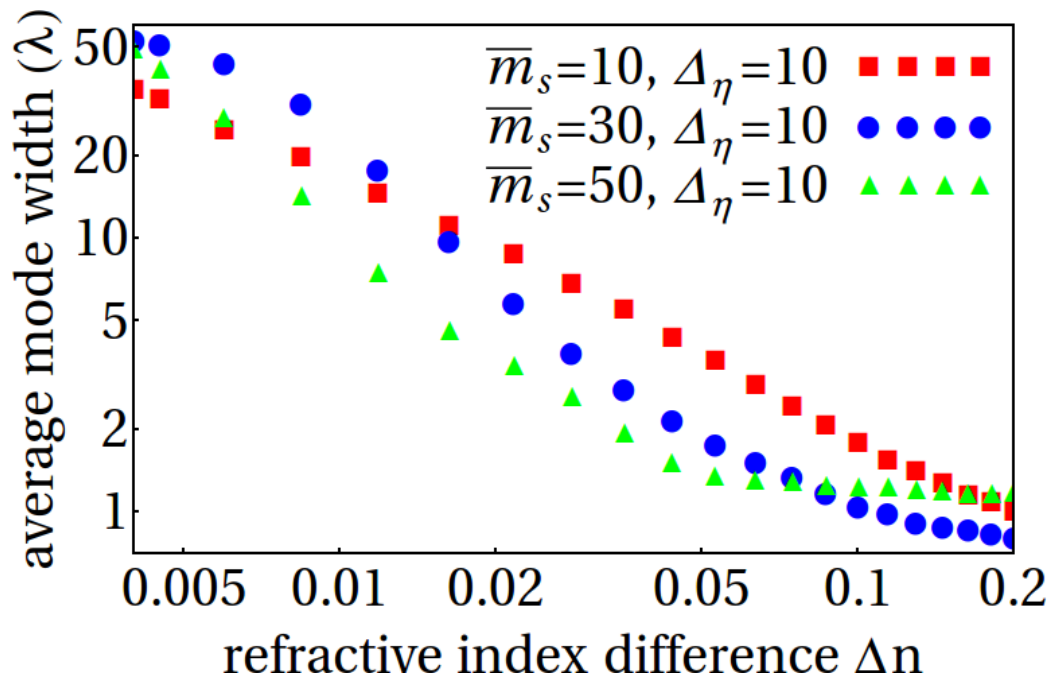

FIG. 10. The average mode widths of different disordered lattices are compared in units of $\lambda$ for $\Delta_\eta = 10$ (fixed level of disorder) as a function of $\Delta n$. The results are presented for three different values of the mean slab width given by $\bar{m}_s = 10$ (squares), $\bar{m}_s = 30$ (circles), and $\bar{m}_s = 50$ (triangles).

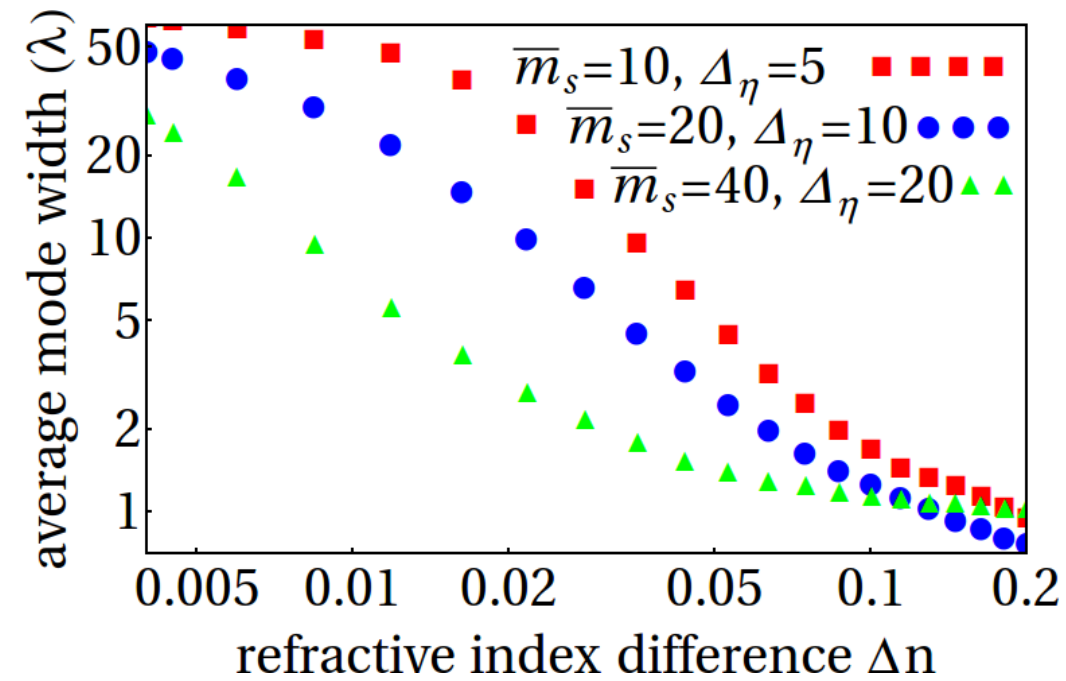

FIG. 11. The average mode widths of different disordered lattices are compared in units of $\lambda$ for $\Delta_\eta = \bar{m}_s/2$ as a function of $\Delta n$. The results are presented for three different values of mean slab width given by $\bar{m}_s = 10$ (squares), $\bar{m}_s = 20$ (circles), and $\bar{m}_s = 40$ (triangles).

### Constant ratio of disorder to the mean slab width

In Fig. 10, we fixed the disorder level at $\Delta_\eta = 10$ and explored the localization behavior for different values of $\bar{m}_s$. In Fig. 11, we carry out a similar study for different values of $\bar{m}_s$, but we choose the disorder level to be half the average lattice thickness, i.e., $\Delta_\eta = \bar{m}_s/2$. We present our results for $\bar{m}_s = 10$ (squares), $\bar{m}_s = 20$ (circles), and $\bar{m}_s = 40$ (triangles). The results from Fig. 11 are quite conclusive and show that over a wide range of $\Delta n$, the lattices with a thicker mean slab width ($\bar{m}_s = 40$) result in a noticeably smaller average mode width. Again, this result is of major practical importance from the fabrication standpoint, because it is easier to construct a disordered lattice from thicker constituent slabs. Therefore, it is important to systematically study, as shown in Fig. 12, the effect of the thickness of constituent slabs on the localization width of the guided modes.

The results presented in Fig. 11 can be used to understand the effect of the optical wavelength on the localization of the optical beam, as well. For a fixed value of $\Delta n$, Fig. 11 indicates that the average mode width (at a fixed wavelength $\lambda$) increases as $\bar{m}_s$ decreases. For a fixed value of the average disorder width, $\bar{d}_s$, wavelength and $\bar{m}_s$ are inversely related, i.e., $\bar{d}_s = \bar{m}_s\Lambda$ ($\Lambda = \lambda/20$); increasing wavelength is associated with decreasing $\bar{m}_s$. Thus, for a given value of the average disorder width, i.e., a given structural disorder, increasing the wavelength (and thus decreasing $\bar{m}_s$) is associated with increasing the average mode width. This observation is consistent with our results presented in the Ref. [19], where we showed that "the shorter incident wavelength results in a smaller localized beam radius". We emphasize that the validity of this reasoning is rooted in the scale invariance which is inherent in the Maxwell's equations [36].

### Optimum average slab width for best localization

In Fig. 12, we plot the average mode width of different disordered lattices as a function of the average thickness of each slab $\bar{d}_s$, for three different cases: $\Delta n = 0.02$ and $\Delta_\eta = \bar{m}_s/2$ (squares); $\Delta n = 0.05$ and $\Delta_\eta = \bar{m}_s/2$ (circles); and $\Delta n = 0.05$ and $\Delta_\eta = 10$ (triangles). The largest average mode widths considered in the three cases belong to $\Delta n = 0.02$ (squares). It is no surprise that the smaller index difference results in weaker localization of the modes. The two different cases with $\Delta n = 0.05$ show similar localization behavior, except at $\bar{d}_s = 0.5\lambda$ ($\bar{m}_s = 10$), where $\Delta_\eta = \bar{m}_s/2 = 5$ (squares) has a clearly lower disorder than for $\Delta_\eta = 10$ (triangles) and results in a weaker localization (larger mode width). For larger values of $\bar{d}_s$, the disorder is sufficiently strong for both cases to make them comparable in the average modal width.

The more interesting feature of all three cases in Fig. 12 is that the localization is weak at small $\bar{d}_s$ and the average mode width rapidly decreases until $\bar{d}_s$ becomes comparable with $\approx 2\lambda$, beyond which the average mode width gradually increases with increasing $\bar{d}_s$. The weak localization at small values of $\bar{d}_s$ can be explained by the fact that when $\bar{d}_s$ is too small compared with the effective "transverse" wavelength, the optical field extends over multiple slabs and simply averages their refractive index in a homogenized fashion. We note that the effective transverse wavelength is defined as $\lambda_T = 2\pi/k_T$, where $k_T$ is the transverse component of the wavevector projected in the direction of the disorder or perpendicular to the direction of propagation. Therefore, for small $\bar{d}_s$, the transverse wavelength can be much larger than $\bar{d}_s$ that it will result in a small scattering cross section. On the other hand, it is clear that very large values of $\bar{d}_s$ cannot result in strong scattering either, because the individual scatterers will be too far apart. Therefore, it should not be surprising that in all three cases considered here, plots of the average mode width versus the average slab width show a minimum value at around $\bar{d}_s \approx 2\lambda$ to $\approx 3\lambda$. As mentioned before, such results are of major practical importance. From Fig. 12, we learn that it is best to keep the feature sizes (average individual slab width) comparable with, or slightly larger than, the wavelength. However, even if the feature sizes are selected to be larger than the optimal value by a factor of two or so, (e.g., due to fabrication constraints), the average mode width does not increase significantly, because the localization is rather forgiving on larger than optimum feature sizes. Moreover, for smaller values of the index difference, the optimum value of $\bar{d}_s$ becomes larger. We note that our observations are in qualitative agreement with our earlier reports regarding the dependence of the localization radius of disordered fibers on the size of the random features [19].

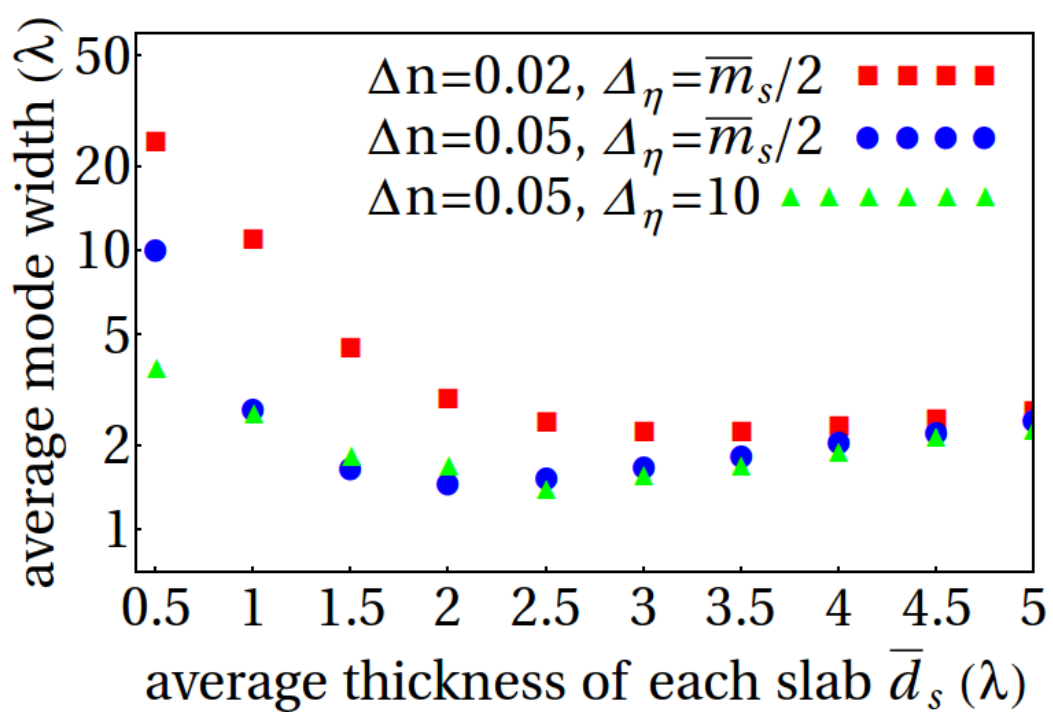

FIG. 12. The average mode widths of different disordered lattices are compared in units of $\lambda$ as a function of $\bar{d}_s$ for three different case: $\Delta n = 0.02$ and $\Delta_\eta = \bar{m}_s/2 = \bar{d}_s/2\Lambda$ (squares); $\Delta n = 0.05$ and $\Delta_\eta = \bar{m}_s/2$ (circles); and $\Delta n = 0.05$ and $\Delta_\eta = 10$ (triangles).

## CONCLUSIONS

The modal language provides a powerful framework to study and analyze transverse Anderson localization of disordered waveguides. In addition, it offers a clear intuitive description of the localization phenomenon independent of the physical properties of the external excitation. Various localization-related phenomena, such as periodic revival of propagating Anderson-localized beams, are easily explained in the modal language. In order to obtain proper Anderson localization, the transverse dimensions of the system must be larger than the localization length; otherwise, boundary effects might disturb or even destroy localization behavior. We also note that the typical picture of an initial expansion and an eventual localization can hold quite well as long as the propagation distance of the beam remains below the revival period.

As expected, stronger localization (smaller average mode width) is obtained if the level of disorder or the refractive index difference between the random constituents is increased. However, the dependence of the average mode width shows a more complex behavior as a function of the average width of the individual random sites in the disordered lattice. In general, weaker localization is observed if the average site width is much smaller or much larger compared with the wavelength of the light. The optimum value for the site width to obtain the maximum localization depends on the design parameters of the disordered lattice, but is usually larger than the wavelength by a factor of two or more.

We anticipate that the modal framework will play an important role in the fundamental understanding, design, and optimization of waveguides and optical fibers that use disorder as their waveguiding mechanism. Extensions of the modal analysis to the 2D disordered systems will be computationally challenging, but are required for proper understanding and optimization of the localization phenomenon in disordered optical fibers; our future efforts will focus in this direction.

## ACKNOWLEDGMENTS

This research is supported by grant number 1029547 from the National Science Foundation.

* mafi@uwm.edu